\documentstyle[preprint,aps]{revtex}

\begin{document}
\title{HBT Interferometry for Sonoluminescence Bubble}
\author{Y. Hama$^1$, T. Kodama$^2$ and Sandra S. Padula$^3$}
\address{$^1$Instituto de F\'{\i}sica, Universidade de S\~{a}o Paulo, \\
Caixa Postal 66318, 05315-970 S\~{a}o Paulo, Brazil\\
$^2$Instituto de F\'{\i}sica, Universidade Federal do Rio de Janeiro, \\
CaixaPostal 68.528, 21945-970 Rio de Janeiro, Brazil\\
$^3$Instituto de F\'{\i}sica Te\'{o}rica, UNESP, Rua Pamplona 145, \\
01405-901 S\~{a}o Paulo-SP, Brazil}
\maketitle

\begin{abstract}
\noindent The two-photon correlation of the light pulse emitted from a
sonoluminescence bubble is discussed. It is shown that several important
information about the mechanism of light emission, such as the time-scale
and the shape of the emission region could be obtained from the HBT
interferometry. We also argue that such a measurement may serve to reject
one of the two currently suggested emission mechanisms, i.e., thermal
process versus dynamical Casimir effect.
\end{abstract}

\bigskip\ 

The sonoluminescence process converts the acoustic energy in a fluid medium
into a short light pulse emitted from the interior of a collapsing small
cavitating gas bubble. Since the discovery of the technique for trapping a
single cavitating bubble by standing acoustic wave\cite{Gaitan,PhysToday},
many remarkable properties have been revealed\cite{Put1,Put2,Put3,Put4}. The
spectrum of emitted light is very wide, extending from the visible to the
ultraviolet regions. The conversion process requires some mechanism of
extraordinary concentration of the energy and is apparently related to
keeping the sphericity of the collapsing bubble while its radius shrinks
more than one order of magnitude. An important fact is that the light
emission takes place within a very short period of time compared to the
typical scale of hydrodynamic motion of the bubble. The sensitivity of the
radiation power and spectrum to physical parameters such as temperature,
pressure, the amplitude of acoustic drive and the gas composition is the
most intriguing aspect of the process and no apparent reason nor any
well-defined systematics are known. In particular, the emission mechanism of
light is still controversial. Some authors attribute the light emission to
the quantum-electrodynamic vacuum property based on the dynamical Casimir
effect\cite{Cas1,Cas2,Cas3}. Others consider that the thermal process\cite
{Put5,Moss,SciAm} like black-body radiation should be the natural
explanation for the process. Non-equilibrium atomic collision process could
also be a strong candidate\cite{Collision}. In any case, the gas dynamics
inside the bubble\cite{Scott,Homol}, in particular the shock wave formation,
seems to play an essential role\cite{Shock}. However, experimental
information on the dynamics of the gas inside the bubble is not available at
present. Due to the short time scale and to the smallness of the emission
region, the precise measurement of these geometric and dynamical information
is quite difficult. Usually, the time dependence of the bubble radius is
measured using the Mie scattering process of laser beams\cite{Put1,Mie}.
Nevertheless, since we have no information on the properties of the gas
during the implosion phase, it is not obvious that the scattered laser
amplitude is really measuring the bubble surface. On the other hand, besides
the knowledge of time elapsed in the process, it would also be desirable to
determine the shape of the emission region. 

\bigskip 

The two-photon interferometry, initially proposed by Hanbury-Brown and Twiss%
\cite{HBT} for measuring stellar sizes, later found its application in the
analysis of high-energy particle production processes\cite{Goldhaber} and is
nowadays widely employed in relativistic heavy-ion collisions\cite
{HBTGeral1,HBTGeral2}. The basic principle is associated to the
Bose-Einstein statistics obeyed by the identical particles involved in the
process and to the chaoticity of the emission mechanism. If the emitting
source has no additional dynamical correlation, then the two-particle
correlation is directly related to the geometrical size of the source.
Several authors\cite{Hama-Sandra,Pratt} have discussed the effect of the
source dynamics on pion interferometry. In this note, we apply this
well-established method to obtain some important information on the dynamics
of the gas inside the bubble while it emits the light pulse. It is
interesting to note that the application of HBT interferometry to the
sonoluminescence bubble ($R\sim 10^{-5}m$) lays just in between the stellar (%
$R\sim 10^{10}m$) and the high-energy physics ($R\sim 10^{-15}m$) scales.

The HBT interferometry method consists in measuring two light quanta in
coincidence. Let $P_2(\vec{k}_1,\vec{k}_2)$ be the probability for
simultaneously detecting two photons with wave-vectors $\vec{k}_1$ and $\vec{%
k}_2$ and $P_1(\vec{k}_i)$ the single-photon probability. The correlation
function $C(\vec{k}_1,\vec{k}_2)$ is then defined as 
\begin{equation}
C(\vec{k}_1,\vec{k}_2)\equiv \frac{P_2(\vec{k}_1,\vec{k}_2)}{P_1(\vec{k}%
_1)\,P_1(\vec{k}_2)}\,\;\;.  \label{c2q}
\end{equation}
If the emitting source is chaotic and static, the correlation function is
simply related to the Fourier transform of its space-time distribution, so
that from the measurement of $C(\vec{k}_1,\vec{k}_2)$ it is possible to
determine the source geometry. In general situations, though, the
relationship between the correlation function and the source geometry is
more complex, being not possible to determine uniquely its geometry only
from the knowledge of $C(\vec{k}_1,\vec{k}_2)$. Nevertheless, once we have a
good guess about the shape and the time development of the emitting source,
we could suggest an appropriate parametrization for the geometry of the
process, and the HBT measurements can be used to determine these parameters.

\bigskip

It should be emphasized that the chaoticity of the emitting source, which
manifests as random phases of emitted signal, plays a crucial role in
relating the correlation function to the geometry of the source. In contrast
to this, if the emission process is coherent (for example, a laser source),
it is well-known\cite{HBTGeral1,Glauber,Wong} that no HBT correlation would
be observed, that is 
\[
C(\vec{k}_1,\vec{k}_2)\equiv 1.
\]
As an immediate consequence, we can distinguish these two extreme scenarios
by a precise HBT measurement. As mentioned before, the mechanisms of light
emission proposed so-far can be classified into two categories: One based on
rather conventional atomic process of thermal origin, and the other due to
the dynamical Casimir effect. In the latter case, a coherent burst of light
would be produced, whereas for the former scenario we expect chaotic
emissions. In both cases the single-photon spectra are similar to that of
the black-body radiation, but an HBT analysis would allow for clearly
differentiating these two emission mechanisms, shedding some light on the
most intriguing feature of the sonoluminescence phenomenon.

\bigskip 

Let us now investigate some examples of non-trivial HBT correlation
functions ($C\ne 1$). For this, we assume that light quanta emitted from
different space-time points has no extra correlation besides that due to the
Bose-Einstein statistics. This is the case when, for example, the emission
process has its origin in atomic collisional or bremsstrahlung mechanisms.
We further assume that the gas inside the bubble be locally in thermal
equilibrium. This does not necessarily mean that the emitted photons are
also in thermal equilibrium with the gas. For simplicity, we here only treat
spherically symmetric sources\cite{Put4}.

Since the time scale of the hydrodynamic motion is considered to be few
orders of magnitude greater than the time scale of the emitting process ($%
\sim 10ps$), we first consider that the source size remains constant during
the light emission. Thus, the time and space dependences is factorized. In
Table 1, we show the analytic expressions of the correlation function for
some typical parametrization of the source density $\rho (r,t)$. The first
three cases (A, B and C) refer, respectively, to a Gaussian type of source,
to a spherical shell and to a homogeneous sphere, all of them having a
Gaussian lapse of time. The fourth example (D) corresponds to an exponential
spatial distribution shining constantly within an interval of time $\tau $.

It could well be possible that the emission region is not static, in which
case the time dependence of the source size should be considered. One may,
for instance, imagine that the radiating source is the spherical domain
behind the expanding shock front formed at the center\cite{Moss,Put5}. In
any case, the fluid velocity is certainly much smaller than the speed of
light so that any dynamical effect due to the fluid motion could be
completely neglected. An emitting source with these characteristics is
represented by the case E. Throughout this study we have taken $\hbar =1$.

\begin{center}
{\bf TABLE 1}
\end{center}

\[
\begin{tabular}{|l|l|l|}
\hline
& Form of the Source & $\,\,\,\,\,\,\,\;\;\;\;\;\;\;\;\;\;\;\;\;\;\;C(\vec{k}%
_1,\vec{k}_2)-1$ \\ \hline
Case A & $e^{-r^2/2R^2}\;e^{-t^2/2\tau ^2}$ & $\,e^{-(\Delta \omega )^2\tau
^2\;}e^{-q^2R^2}/2$ \\ \hline
Case B & $\delta (r-R)\;e^{-t^2/2\tau ^2}$ & $\,e^{-(\Delta \omega )^2\tau
^2\,}[{\rm \sin }(qR)/(qR)]^2/2\,$ \\ \hline
Case C & $\Theta (R-r)\;e^{-t^2/2\tau ^2}$ & $\,9\,e^{-(\Delta \omega
)^2\tau ^2}\{\,[{\rm \cos }(qR)-{\rm \sin }(qR)/(qR)]/(qR)^2\}^2/2$ \\ \hline
Case D & $e^{-r/R}\;\Theta (3\tau ^2-t^2)$ & $\left[ \sin (\Delta \omega 
\sqrt{3}\,\tau )/\Delta \omega \sqrt{3}\,\tau \right] ^2\;\left(
1+q^2R^2\right) ^{-4}/2$ \\ \hline
Case E & $\Theta (\dot{R}\,t-r)\,e^{-t^2/\tau ^2}\Theta (t)$ & $9\left|
I\right| ^2/(8\mu ^6),$ \\ 
&  & $I=-i\sqrt{\pi }\left[ (1+\mu z^{+})W(z^{+})-(1-\mu
z^{-})W(z^{-})\right] -2\mu $ \\ \hline
\end{tabular}
\]
Here, $q=\left| \vec{k}_1-\vec{k}_2\right| \,$, $\Delta \omega =\omega
_1-\omega _2=c(k_1-k_2),$ $\mu =\dot{R}\,\tau \,q,$ $z^{\pm }=(\Delta \omega
\pm \dot{R}\,q)\tau /2$ and $\,W(z)\equiv e^{-z^2}{\rm erfc}\,(-iz)$. $R$ is
the spatial extension parameter and $\tau $ is the time span parameter of
the source and $\dot{R}$ is the velocity of the shock wave. The case A was
studied by Trentalange and Pandey \cite{Trenta}, following the prescription
given in \cite{Neu}. However, as pointed out by Slotta and Heinz\cite{Heinz}%
, the current conservation should be correctly taken into account, which
results in a slight deviation from eq.(6) of Ref.\cite{Trenta}. Note that in
the first four cases (A-D) above, the correlation function are written in
the form 
\begin{equation}
C(\vec{k}_1,\vec{k}_2)=1+\frac 12T(\Delta \omega )\Phi (q)\,\;\;,
\label{C(k1k2)}
\end{equation}
which is a consequence of factorized form of the source. Therefore, to
determine the time span parameter $\tau $, it is convenient to plot the data
for fixed $q$. If the source is really factorized in space and time, then
plots of $\log (C-1)$ vs. $\left( \Delta \omega \right) ^2$ for different
pairs of $(\vec{k}_1,\vec{k}_2)$ should generate a set of identical curves
just shifted from each other for different values of $q=\left| \vec{k}_1-%
\vec{k}_2\right| $. In Fig.1, we show examples of such plots, with $\tau
=1ps,\,R=1\mu m$ in the blue-light domain $(k_1\simeq k_2\simeq 4\times
10^7\;m^{-1})$. The solid lines refer to the case A, the filled circles to
the case D and dotted lines to the case E. As can be seen in this figure, if
the time span is Gaussian and factorized from the space dependence, then
these curves should be parallel straight lines whose slopes would give the
value of the parameter $\tau $. If the time span is not Gaussian, then the
curves are not straight lines, but similar to those for the case D. However,
even in this case, the first derivatives of the corresponding curves with
respect to $(\Delta \omega )^2$ at the origin provide an estimate of the
parameter $\tau $. The information about the source spatial distribution, $%
\Phi (q)$, is determined by means of the intersection of these lines with
the abscissa. For the case E of Fig.1, the velocity of expansion was taken
to be $\dot{R}=2\times 10^{-4}c$, where $c$ is the speed of light. If the
velocity is smaller than this value, the lines corresponding to different $q$%
's come closer to one another. This happens because, for smaller velocities,
the effective emission region becomes smaller and, consequently, the
correlation function becomes broader and slow-varying with $q$. Note that
for the case E, the lines are not parallel, reflecting the non-factorized
emission source.

In practice, such an analysis might be limited by experimental conditions.
In particular, to get a meaningful result for the $\tau $ parameter, $\Delta
\omega $ should be measured within a resolution of the order of $1/ps$,
which corresponds to measuring the photon energy itself within precision of $%
\delta \omega /\omega <10^{-4}$. This might not be easily achieved. However,
even if the energy resolution is not high enough, the spatial factor in eq. (%
\ref{C(k1k2)}) can independently be analyzed if the source is factorized.
Being so, the HBT correlation function of two photons of approximately the
same energy, under a poor energy resolution (within $\delta \omega $), would
lead to an angular correlation given by 
\[
C(\vec{k}_1,\vec{k}_2)\rightarrow C(\vec{k}_1,\vec{k}_2)=1+\frac 12\langle
T(\delta \omega \rangle \,\Phi (q)\,.
\]
Since the multiplicative factor $\langle T(\delta \omega )\rangle $ is
independent of $q$, it can easily be eliminated from $C$ near the origin, $%
q=0$. The necessity of the above renormalization of the correlation function
at the origin as a consequence of a poor energy resolution of HBT
measurements has been pointed out in \cite{Hama-Sandra}. The function $\Phi
(q)$ in (\ref{C(k1k2)}) behaves quadratically in $q$ around $q=0$. Namely, 
\[
\Phi (q)=\left\{ 
\begin{array}{c}
1-R^2q^2+\;\cdots ,\;\;\;\;\;\text{Gaussian (A),} \\ 
1-R^2q^2/3+\cdots ,\;\;\;\;\;\;\text{Shell (B),} \\ 
1-R^2q^2/5+\cdots ,\;\;\;\;\;\;\;\;\text{Sphere (C),} \\ 
\;1-4R^2q^2+\cdots ,\;\;\text{Exponential (D),}
\end{array}
\right. 
\]
so that we can determine the parameter $R$ by the curvature of $\Phi (q)$ at
the origin. By defining 
\[
\kappa \equiv \left. -\frac{d^2\Phi }{dq^2}\right| _{q=0},
\]
we have 
\[
R=\left\{ 
\begin{array}{c}
\sqrt{\kappa /2},\;\;\;\;\;\text{Gaussian (A),} \\ 
\sqrt{3\kappa /2},\;\;\;\;\;\;\;\text{Shell (B),} \\ 
\sqrt{5\kappa /2},\;\;\;\;\;\;\;\;\;\text{Sphere (C),} \\ 
\sqrt{\kappa /8},\;\;\text{Exponential (D).}
\end{array}
\right. 
\]
To distinguish among the shapes of the correlation functions reflecting
different source density distributions, we would need to know the behavior
of $\Phi (q)$ in a wider range in $q$. To stress the differences among the
four cases above, let us introduce the variable $X$ defined as 
\[
X\equiv \sqrt{\kappa /2}\;q\,.
\]
Then, by definition, $\Phi $ behaves as 
\[
\Phi =1-X^2+\cdots \,,
\]
near $X=0$. In Fig.2, we compare the behavior of $\Phi $ as a function of $X$
for the four cases. The continuous line corresponds to the Gaussian spatial
distribution, the triangles to the sphere, the squares to the spherical
shell and, finally, the circles correspond to the exponential density
distribution. As can be seen in this figure, the differences among these
curves are not striking near the origin, but if the data are precise enough (%
$\sim 3$ order of magnitudes) in a sufficiently wide range of $X$, we may
determine the shape of the source function $\rho $.

\bigskip 

As already mentioned, if the emitting process were coherent, the behavior of
the correlation function would be entirely different from the ones just
discussed. As anticipated in the beginning, a precise measurement of
two-photon correlation function of the light quanta emitted by a
sonoluminescence bubble would allow for distinguishing between chaotic and
coherent emission mechanisms. However, we should stress that, even in the
case of a chaotic source, a poor energy-resolution experiment could lead to
a result similar to the one expected in the case of a coherent source, due
to the factor ${\rm exp}\,(-\tau ^2\Delta \omega ^2)$. Therefore, a very
high energy-resolution experiment is required to clearly differentiate
between these two opposite scenarios.

\bigskip 

This work has been partially supported by FAPESP (Contract 95/4635-0), CNPq
and FAPERJ. One of us, (T.K.) expresses his gratitude to J. Rafelski,
I.Scott and H-T. Elze for stimulating discussion on sonoluminescence
phenomena and encouragements. Discussions with C.E. Aguiar, L. Pimentel and
P.A.Nussenzweig are gratefully acknowledged.

\newpage\ 

\begin{center}
{\bf FIGURE\ CAPTIONS}
\end{center}

\begin{description}
\item[FIG.1]  Correlation functions plotted as functions of $\Delta \omega $
for varios values of $q$. The solid lines refer to the case A, filled
circles are for the case D and dotted lines are for the case E. The first
two cases are examples of factorized sources, whereas the lines for the case
E are not parallel.

\item[FIG.2]  Geometrical form factor $\Phi (q)$ plotted as functions of $X=%
\sqrt{-\frac 12\frac{d^2\Phi (0)}{dq^2}}q$. The solid curve corresponds to
the Gaussian source, the triangles to the spherical source, the squares to
the spherical shell and the circles to the exponential distribution.
\end{description}

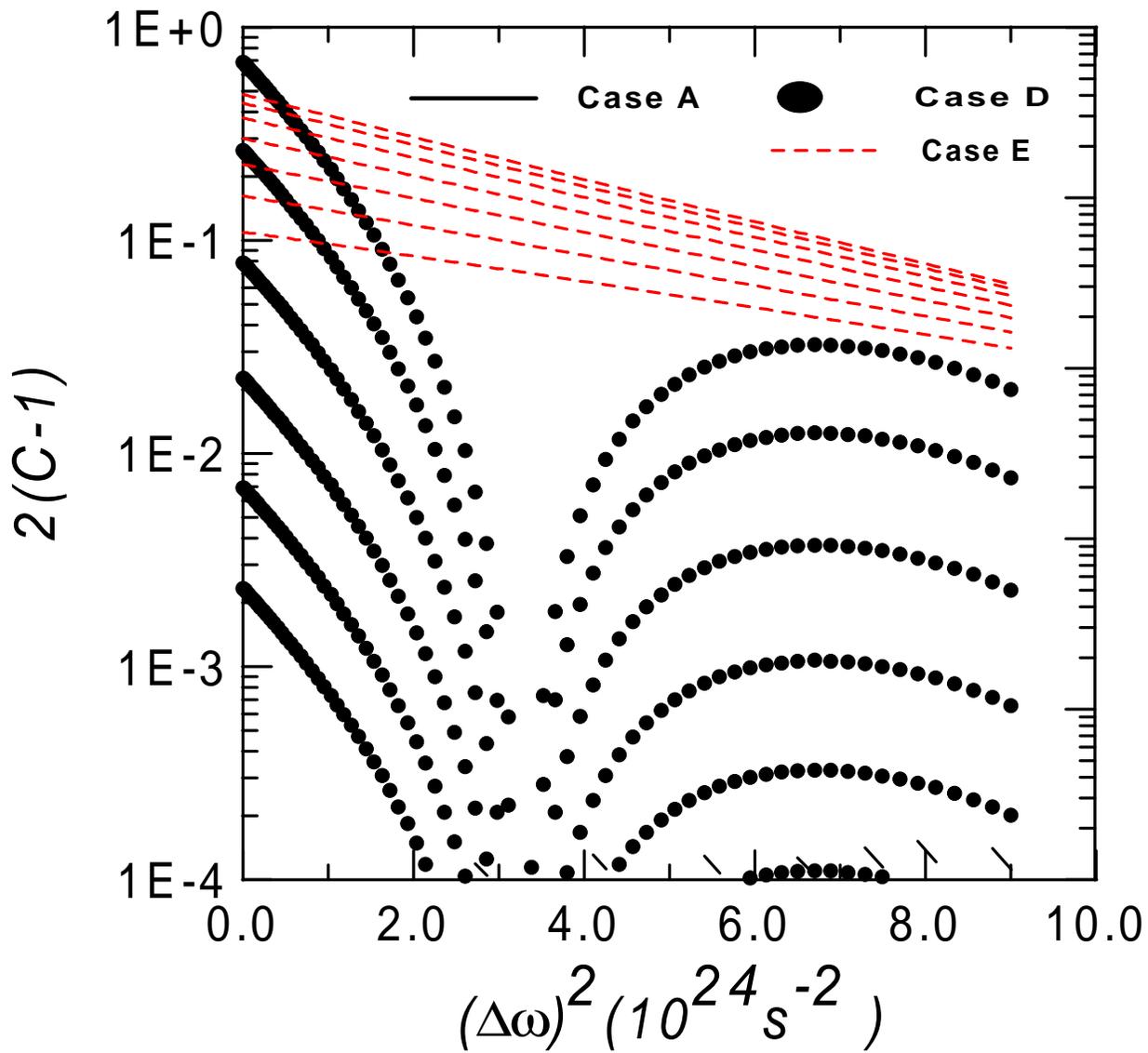
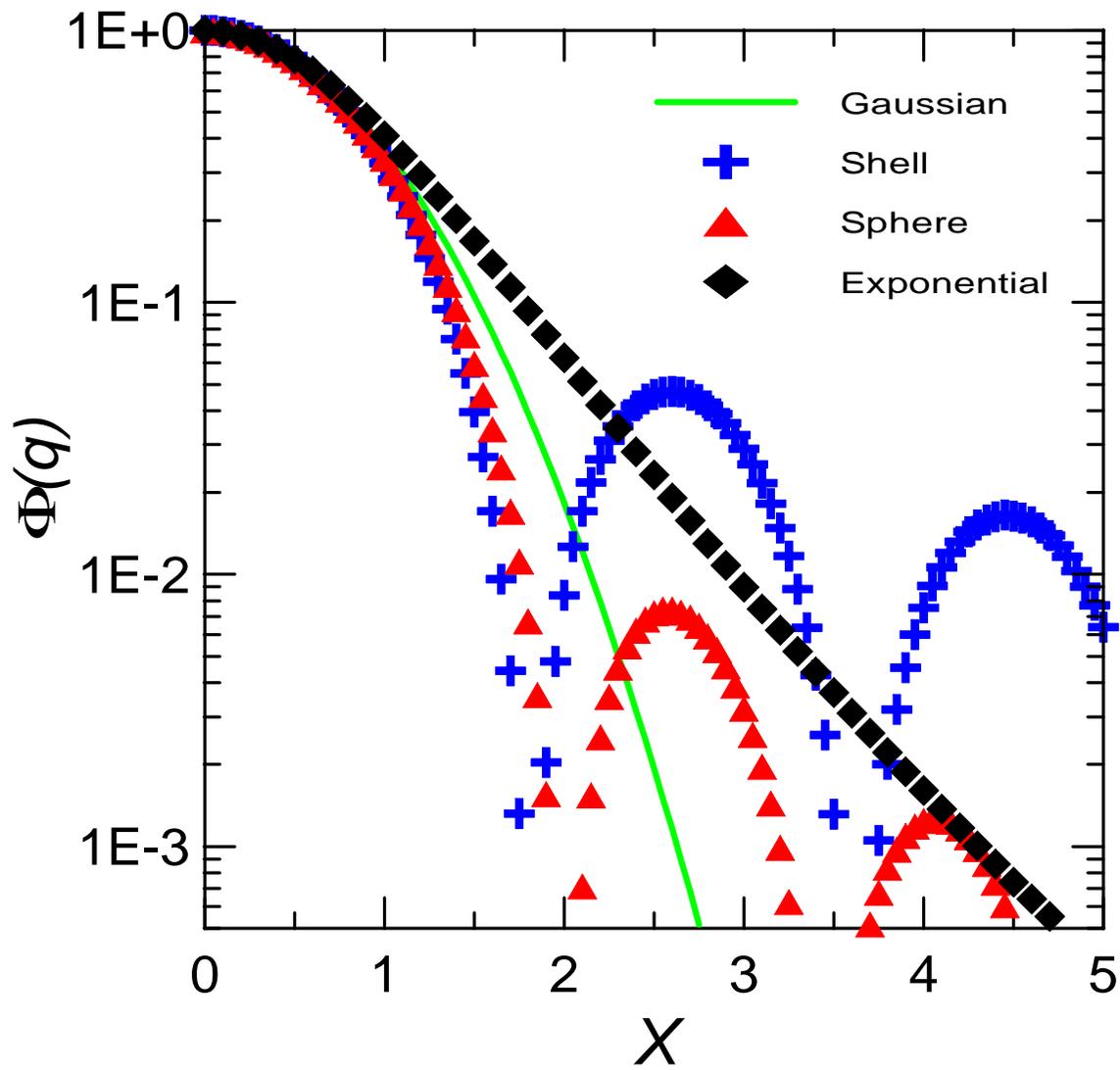
\end{document}